\documentclass[aps,prb,floatfix,showkeys,showpacs]{revtex4}
\usepackage{graphicx}
\usepackage{dcolumn}
\usepackage{bm}
\usepackage{amsmath}

\begin{document}

  \title{Exciton-phonon scattering and photo-excitation dynamics in J-aggregate microcavities.}
  \author{Paolo~Michetti}
  \affiliation{Dipartimento di Fisica, Universit\`a di Pisa,
    Largo Bruno Pontecorvo 3, 56127 Pisa, Italy}
  \email{michetti@df.unipi.it}

  \author{Giuseppe C.~La Rocca}
  \affiliation{Scuola Normale Superiore and CNISM
    Piazza dei Cavalieri 7, 56126 Pisa, Italy}


  \begin{abstract}
    We have developed a model accounting for the photo-excitation dynamics and
    the photoluminescence of strongly coupled J-aggregate microcavities.
    Our model is based on a description of the J-aggregate film as a disordered Frenkel exciton
    system in which relaxation occurs due to the presence of a thermal
    bath of molecular vibrations.
    In a strongly coupled microcavity exciton-polaritons are formed, mixing superradiant excitons and cavity photons.
    The calculation of the microcavity steady-state photoluminescence, following a
    CW non resonant pumping, is carried out.
    The experimental photoluminescence intensity ratio between upper and lower
    polariton branches is accurately reproduced.
    In particular both thermal activation of the
    photoluminescence intensity ratio and its Rabi splitting
    dependence are a consequence of the bottleneck in the relaxation, occurring at the
    bottom of the excitonic reservoir.
    The effects due to radiative channels of decay of excitons and to
    the presence of a paritticular set of discrete optical molecular vibrations
    active in relaxation processes are investigared.   
  \end{abstract}

  \maketitle

  \section{\label{sec:level1}Introduction}
  Strong coupling in solid state microcavities (MCs), has been
  exhaustively studied in inorganic quantum well devices \cite{book}.
  In these structures Wannier excitons and photon cavity modes are mixed
  in coherent excitations, called polaritons.
  A number of interesting phenomena, due to the
  mixed nature of such eigenstates, such as the
  parametric processes \cite{parametric} and non-equilibrium polariton
  condensation \cite{condensation}, were revealed.

  On the other hand, organic strongly coupled MCs have been developed since 1998 \cite{lidzey1}, by using different
  kinds of optically active organic layers, among which the cyanine
  dye J-aggregate films are probably the most typical.
  Organic materials possess Frenkel excitons,
  instead of Wannier ones, with large binding energy and oscillator strength.
  Because of this, organic MCs allow obtaining values of the
  Rabi splitting up to $300$ meV \cite{hobson} and offer the possibility
  of easily observing polaritons at room temperature \cite{lidzey2}.
  Electroluminescence was demonstrated in a J-aggregate strongly
  coupled microcavity LED \cite{led} at room temperature.
  Peculiar molecular phenomena, like the observation of strongly
  coupled vibronic replicas have also been demonstrated \cite{holmes,fontanesi}.
  A detailed review about organic MCs can be found in Refs.\cite{forrest,lidzeyrev}

  The nature of polaritons in disordered organic MCs has been addressed from the theoretical point of view
  \cite{agranovich,michetti,gartstein,michetti1}, showing the coexistence of
  delocalized and partially localized polaritons.
  The possible interaction with molecular vibrations was discussed
  \cite{litinskaia, michetti2}, the effect of anisotropy in organic
  crystal was characterized \cite{zoubi1}, as well as the possible occurrence of
  non-linear phenomena \cite{zoubi2} for high density of polaritons.

  We believe that, as in the case of inorganic MCs, theoretical
  efforts towards a qualitative, but also quantitative, understanding of the
  relaxation processes in organic MCs dynamics are an essential
  support to the work of experimental teams involved in the study of organic microcavities,
  in principle able to give rise to a rich variety of phenomena, similar
  to what obtained with their inorganic counterpart and more, thanks to
  the infinite possibilities of organic chemistry.
  A recent model described the dynamics of a J-aggregate MC due to the strong coupling of
  polariton with optical
  molecular vibrations \cite{chovan} thorough the application of a quantum kinetic theory.
  Our approach \cite{michetti2}, even though restricted to a rate equation description of the relaxation
  processes, allows for a more
  flexible and complete description of the physical system at hand, accounting for the
  photoluminescence of a 2D
  MC starting from an accurate description of the bare active
  material (the J-aggregate film).
  Besides we are also able to perform time dependent simulations of
  J-aggregate microcavity photoluminescence \cite{michettime} and 
  estimate their radiative decay time in linear pumping regime. 
  In particular, we here extend our model, which includes the presence of a large number
  of weakly coupled excitons, considering the occurrence of a radiative relaxation
  channel from such excitons to the cavity polaritons, as first suggested in ref.\cite{lidzeyA}.
  We also include the effects on the polariton relaxation due to the
  presence of a discrete spectrum of molecular vibrations, as in
  \cite{chovan}, which however are here supposed to scatter weakly
  with the electronic excitations without taking part directly in
  the cavity-polariton formation.

  \section{The model}
   A J-aggregate can be thought as a chain of $N_d$ dye molecules linked by
  electrostatic interactions.
  Its excitations are described by the Frenkel Hamiltonian:
  \begin{equation}
    H=\sum_{i}^{N_d} E_{i} b_{i}^\dag b_{i} +\sum_{i\neq
      j}^{N_d}V_{i,j}(b_{i}^\dag b_{j}+ b_{j}^{\dag}b_{i})
    \label{eq:HFrenkel}
  \end{equation}
  where $b_i^\dag$ are the monomer exciton creation operators, corresponding to
  Gaussian distributed energies $E_{i}$, with standard deviation
  $\sigma$, to account for static disorder.
  The hopping term is given by $V_{i,j}=-J/|i-j|^{3}$
  where $J>0$ is the nearest neighbor coupling strength.
  The eigenstates are found by direct diagonalization of the Frenkel
  Hamiltonian and are described by the operators $ B_{\alpha}^{\dag}=\sum_{i}c_{i}^{(\alpha)} b_{i}^{\dag}$,
  where $c_{i}^{\alpha}$ is the coefficient of the $\alpha$-th  exciton
  on the i-th monomer.
  The oscillator strength of each state is given by
  $F_{\alpha}=|\sum_{i}c_{i}^{(\alpha)}|^{2}$,
  with $\sum_\alpha F_\alpha =N_d$, having the molecules equal
  transition dipole and being the wavelength much greater than the aggregate length.
  The excitations are delocalized Frenkel excitons with an energy
  dispersion curve, called the J band, where most part of the
  aggregate oscillator strength is concentrated on the superradiant
  bottom states \cite{kobayashi}.
  Higher energy excitons very weakly couple with light.
  While the hopping term determines the formation of the J band,
  static disorder gives rise to a low energy tail in the exciton DOS,
  corresponding to partially localized, inhomogeneously
  broadened, superradiant eigenstates, responsible of the film
  luminescence and, in MCs, of the strong light-matter coupling.

  Delocalized polaritons are the result of the strong light-matter
  interaction between such superradiant excitons, localized on
  each aggregate, and the photon cavity modes, ideally extended over
  the whole structure.
  To treat the MC case we include a model
  polariton wavefunction of wavevector $k$, linking the J aggregate excitons into the
  polariton states \cite{michetti2}.
  The total exciton and photon fraction, $C_{k}^{(ex)}, C_{k}^{(ph)}$, and the
  polariton dispersion curve $E_{U,L}(k)$ are given by the usual two coupled
  oscillator model.
  The modulus of the exciton coefficients $\phi_{\alpha,I}^{(k)}$, describing
  how each exciton $\alpha$ of $I$-th aggregate
  participates in the polariton states, are the only relevant quantities
  entering our model.
  We consider the polaritons to be extended with equal degree on
  each one of the $N_{agg}$ aggregates and make the assumption that
  the weight of each exciton in the polariton states
  is proportional to its oscillator strength, so that
  $|\phi_{\alpha,I}^{(k)}|=\sqrt{F_{\alpha}/N_d}$.

\begin{figure}
    \centering
    \includegraphics[width= 8.3cm]{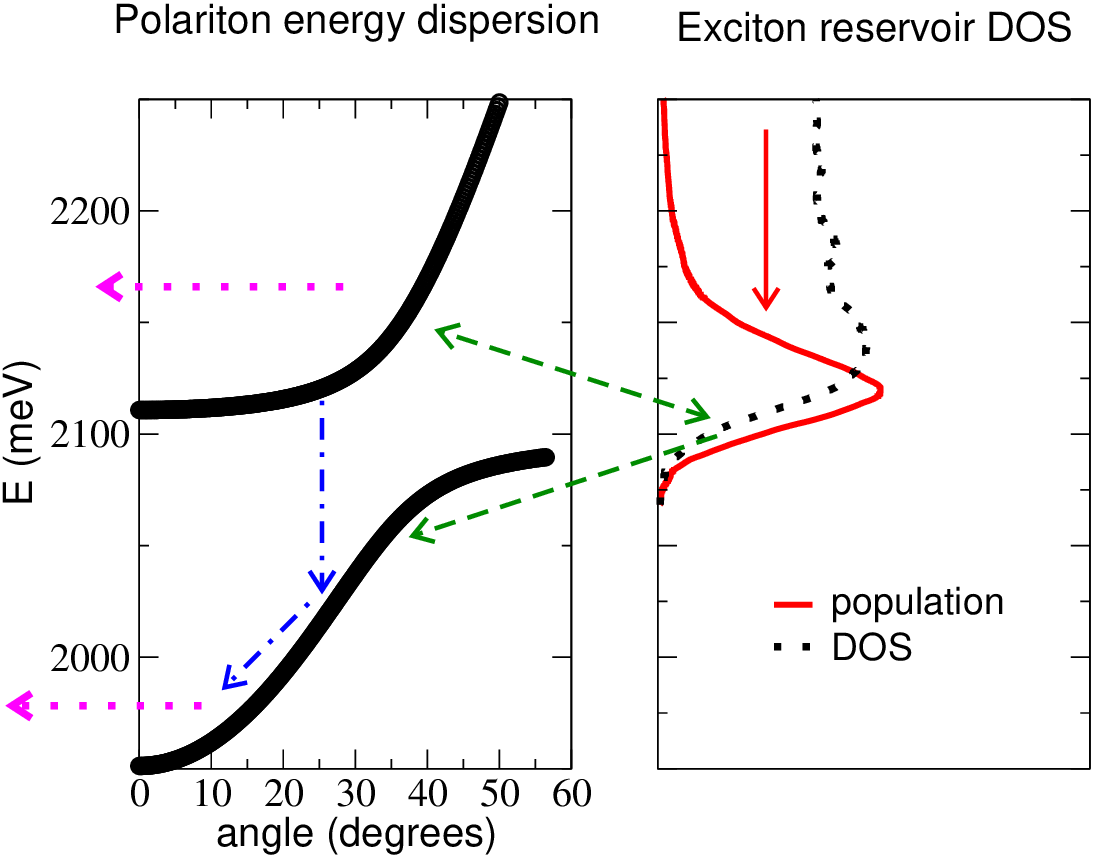}
    \caption{(Color online) On the left polariton dispersion curves for
      $\Delta=80$~meV, on the right the exciton reservoir DOS (dotted line) and
      steady state population (countinuous line) are represented.
      The possible scattering mechanisms are sketched out: exciton to
      exciton (continuous arrow), exciton to/from polariton (dashed
      arrow), polariton to polariton (dotted-dashed arrow), polariton
      radiative decay (dotted arrow).
    }
    \label{fig:schema}
  \end{figure}

  In organic microcavities the strong coupling region, where
  polaritons do form, is typically extended till about $q_{max} = 10^5$
  cm$^{-1}$, where the Bragg mirrors stop-band starts to fail, or is
  also more restricted by inhomogeneous disorder \cite{agranovich,
  michetti1}.
  Therefore the excitation spectrum is composed, besides polaritons, of a number of
  uncoupled molecular excitons that form an excitonic
  reservoir (ER) \cite{agranovich, litinskaia, michetti1} as shown in Fig.\ref{fig:schema}.
  The excitonic part of the polaritons are formed by a Bloch sum, with
  $q<q_{max}$, made from the superradiant molecular excitons of the
  J-aggregates, while the ER is composed by uncoupled superradiant excitons from the remaining part of the
  Brillouin zone, plus a number of higher energy  ``dark'' excitons.
  We note that the ratio between polaritons and uncoupled excitons of
  the ER can be of the order of $ (q_{max}*R)^2/N_d \approx 10^{-3}$ or less.

  The dynamics of the system is described by rate
  equations for the population $f_{i}(t)$ of the state $i$, including the scattering rates, due to a weak linear
  exciton-phonon coupling, among all the kinds of particles present in
  the system which have a non-null exciton component.
  Dealing with a 2D microcavity, we assume isotropic conditions,
  in which the polariton population only depends on the modulus of the
  wavevector, in order to reduce the dimensionality of the problem.
  We consider localized vibrations, that can be due to the molecules
  itself or to the local environment, located on each chain site with
  a continuous spectrum, choice that follows literature reports \cite{bednarz,heijs} on
  J-aggregate simulations.
  We will address this model as the continuous spectum of vibration
  (CSV) model.
  We also extend the model to account for a series of discrete optical
  vibrations in what we will refer as the discrete vibrations model (DV
  model), such specific modes can be of particular relevance when their energy is nearly resonant
  with the Rabi splitting \cite{lidzeyA, chovan}.
  In particular the DV model includes a continuum of vibrations for energy below $25$~meV and
  two discrete molecular vibrations of $E_1=40$~meV and $E_2=75$~meV, with an
  exciton-phonon coupling factor $g_1=0.4$ and $g_2=0.5$, respectively.
  The values of the molecular vibrations are chosen following the Raman
  experiments on J-aggregate films, as reported in \cite{chovan}.

  The scattering rates from the state $i$ to the $i'$, which can either
  be an exciton $i=n=(\alpha,I)$, the $\alpha$-th  exciton of the $I$-th aggregate, or a
  delocalized polariton of wavevector $k$, can be expressed in an
  unique compact notation.
  If we consider the interaction with a
  continuous spectrum of vibrations, for which the vibrational DOS is
  non-zero for any energy difference $\Delta
  E= E_{i'}-E_i$ between final and initial state, we adopt the following expression
  \begin{eqnarray}
    W_{i',i} = W_{0} \Xi_{i,i'}
    \times (N_{|\Delta E|}+\Theta_{(\Delta E)}) \Big(\frac{|\Delta E|}{J}\Big)^{p}
    \label{eq:Wcont}
  \end{eqnarray}
  where $W_0$ and $p$ are parameters related to the
  exciton-photon interaction energy and to the shape of the
  vibration spectrum, $N_{|\Delta E|}$ is the Bose-Einstein occupation
  factor, $\Theta_{(\Delta E)}$ the step function.
  If we instead consider the interaction with a molecular vibration of
  discrete energy $E_{vib}$ we obtain
  \begin{eqnarray}
    W_{i',i} = \frac{2 \pi g^2 E_{vib}^2}{\hbar} \Xi_{i,i'}
    \times (N_{|\Delta E|}+\Theta_{(\Delta E)}) \delta(|\Delta E|-E_{vib})
    \label{eq:Wdisc}
  \end{eqnarray}
  where $\delta(E)$ is the Dirac delta and $g$ the exciton-phonon
  coupling parameter for the vibration of energy $E_{vib}$.
  What distinguish the different processes is the overlap
  factor $\Xi_{i,i'}$.
  In fact the emission or absorption of localized vibrations promotes the
  scattering among states that have an exciton participation on
  the same molecular site.
  Therefore the factor $\Xi_{i,i'}$ take in account of the partial
  exciton nature of polaritons, of their delocalized nature and are
  proportional to an excitonic overlap factor $I$ obtained by carrying out
  the sum of the excitonic participation of the initial and final state on each
  molecular site.  
  The overlap factor is one of the following:
  \begin{eqnarray}
    \Xi_{n,n'} &=& \delta_{I,I'} I_{\alpha,\alpha'}
  \end{eqnarray}
  \begin{eqnarray}
   \Xi_{k,n'} &=& \frac{D(k)|C_k^{(ex)}|^2}{N_{agg}}
    I_{n'}^{(ex-pol)}\\
    \Xi_{n',k} &=& \frac{|C_k^{(ex)}|^2}{N_{agg}} I_{n'}^{(ex-pol)}\\
    \Xi_{k',k} &=& \frac{D(k')|C_{k'}^{(ex)}|^2|C_k^{(ex)}|^2}{N_{agg}^2} I^{(p-p)}
  \end{eqnarray}
  respectively, in the case of scattering from an exciton to another
  exciton ($i=n$, $i'=n'$), where $I_{\alpha,\alpha'}$ is the excitonic
  overlap factor;
  or for the scattering from an exciton $n'$ to a polariton $k$, with
  $I_{n'}^{ex-pol}$ their excitonic overlap factor, and $D(k)$ the number of states corresponding, in the
  inverse space grid, to the polariton variable of modulus wavevector
  $k$; or when describing the opposite process ($i=k$, $i'=n'$); or
  in the latter case of scattering between two polaritons, where
  $I^{(p-p)}$ is an excitonic overlap factor between polaritons (the derivation of the rates can be found in Ref. \cite{michetti2}).

  We also add a radiative channel of decay from the ER to the
  polaritons, accounting for the presence in the MC of a fraction of excitons weakly
  coupled with light.
  In fact, inside a strongly coupled cavity a fraction of
  weakly coupled excitons can decay radiatively pumping the photonic
  part of polaritons \cite{lidzeyA}.  
  We model the spontaneous emission of polaritons from the ER similar
  to the bare film photoluminescence  times a switching
  factor $\beta$:
  \begin{equation}
    W_{k,\alpha}^{rad}= \beta \gamma_{\alpha}\frac{D(k)|C_{k}^{(ph)}|^{2}g_{\alpha}(E_{k})}{\sum_{k'}D(k')|C_{k'}^{(ph)}|^{2}g_{\alpha}(E_{k'})},
    \label{eq:betarad}
  \end{equation}
  the contribution of the $\alpha$ excitons is proportional to the bare exciton luminescence
  ratio $\gamma_\alpha$, the transfer matches energy conservation, while the
  normalization is chosen to obtain a net exciton radiative decay
  towards polaritons of $\beta$ times the spontaneous
  emission ($\gamma_{\alpha}$). 
  $g_{\alpha}$ is a Lorentzian shape broadening function, where the
  energy uncertainty is given by the finite particle lifetime.

  The damping rate of polaritons, due to photon escape through the cavity mirrors, is given
  by $\Gamma_{i}= |C_{i}^{(ph)}|^{2}/\tau$, where typically
  $\tau=35$ fs.
  For the upper and lower polaritons we consider the reciprocal space
  up to $q_{max}=9\times 10^4$ $cm^{-1}$, discretized in a grid of $280$ points.
  The effect of the inhomogeneous disorder are taken in account
  averaging the scattering rates over an ensemble of $10^3$
  J-aggregates each having $N_d=100$ monomers.
  The dyes hopping energy is fixed at $J = 75$ meV, and the dye spontaneous
  emission lifetime is $\gamma_0= 0.33$ ns$^{-1}$;
  while the parameters $\sigma=0.54J$, $p=0.8$ and $W_0=3.2J/\hbar$,
  have been fixed in order to obtain the best fit the bare J-aggregate film
  absorption and emission features \cite{michetti2}.

  \section{Photoluminescence simulation}
  \subsection{Continuous spectrum of vibrations}
  Solving the rate equations for the J-aggregate microcavity, with the
  same procedure described in ref.\cite{michetti2}, we obtain
  the polariton steady state population, from which we calculate the
  angle resolved photoluminescence, by imposing the conservation of energy
  and in-plane momentum between cavity-polaritons and external
  photons. 
  For both the CSV and the DV models, we draw the indication that the
  major part of the population is frozen on the ER.
  The origin of this bottleneck is due essentially to three reasons: the small number of polariton
  states with respect to the ER ones, second the interaction between a
  localized vibration and a polariton is reduced by the delocalized,
  and mixed, nature of exciton-polaritons, in fact the scattering rates scale
  as $1/N_{coher}$, where $N_{coher}$ is the number molecular sites
  over which the polariton is delocalized.
  Third is the small photon lifetime, due to the low quality factor of
  organic MCs.
  The photoluminescence can be thought of as the result of
  the escape through the cavity mirrors of a particle, following the
  previous scattering on the polariton branches from the ER, with 
  the absorption or emission of a vibrational quantum.

  In Fig.\ref{fig:schema} we draw a sketch of the excited states on a
  J-aggregate microcavity and of the possible scattering
  processes following non resonant pumping.
  In particular we stress that the relaxation inside a J-aggregate is faster than ER to
  polariton scattering and leads to a steady state population (shown
  in the right figure in continuous line), from which polariton
  pumping occurs.
  The ER DOS is inhomogeneously broadened by static disorder and ER
  steady state population is similar to that of a non-cavity sample.
  We note here that the average difference between the energy of the steady state
  population and the film absorption maximum results to
  be about $\varepsilon=15$~meV.

  \begin{figure}
    \centering
    \includegraphics[width= 8.3cm]{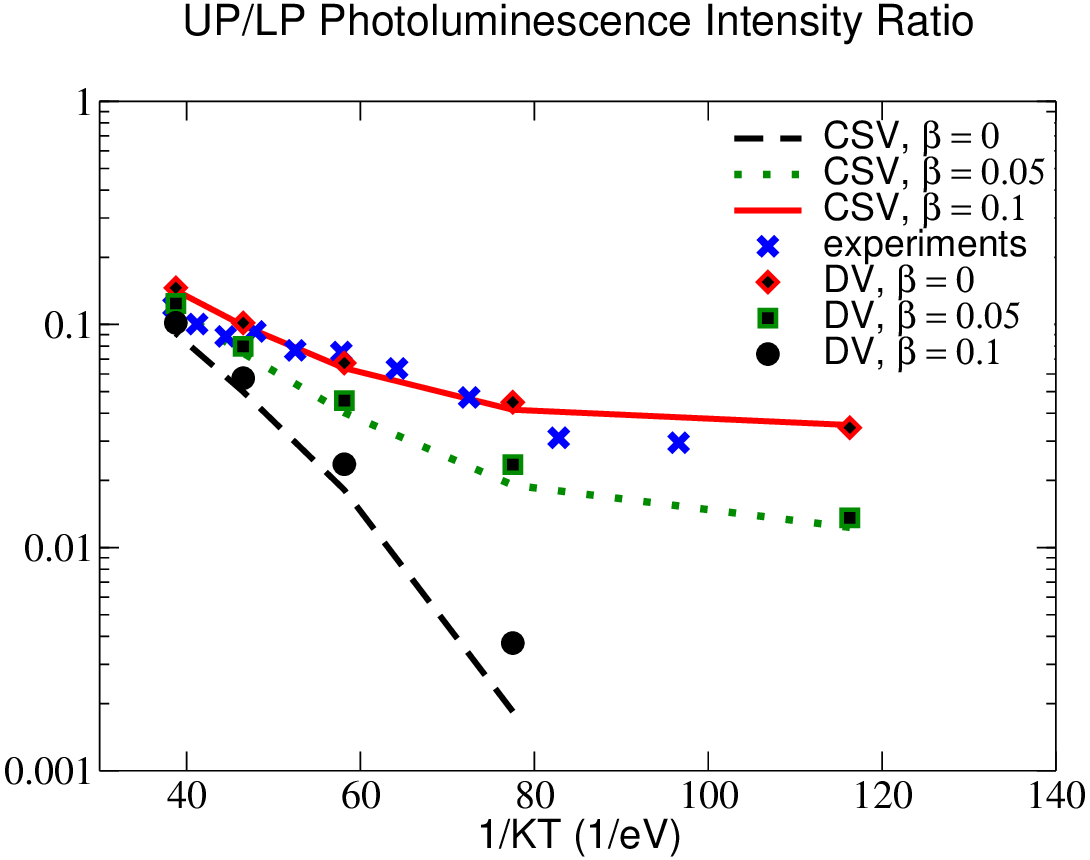}
    \caption{(Color online) The $C$ ratio as a function of temperature for a MC with
      $\Delta=130$~meV.
      The experimental results from Ref.\cite{ceccarelli} (X marks)
      are compared with simulations with increasing importance of the radiative channel $\beta^{rad}=0$, $0.05$, and
    $0.1$, for both the model with continuous spectrum of vibrations and the
    model DV model.}
    \label{fig:temp}
  \end{figure}
  An experimentally accessible parameter, used to analyze
  the effectiveness of relaxation in organic microcavities,
  is the photoluminescence intensity ratio $C= UPPL/LPPL$ recorded at resonance
  angle, where polariton branches are half matter and half light, and their energy distance from the
  exciton resonance is half the Rabi splitting ($\Delta/2$).
  We focus on the temperature dependence of the $C$
  ratio between $50$ and $300$~K for a MC with a Rabi splitting of
  $\Delta=130$~meV, shown in Fig.\ref{fig:temp}, and compare it with the experimental result reported by D.G. Lidzey's  group \cite{ceccarelli}.
  With $\beta^{rad}=0$ the $C$ ratio shows a smooth thermal activation, but it is
  difficult to get any UP PL at low temperature.
  This is consistent with the fact that UP PL follow the absorption of
  a vibrational quantum from a particle at the bottom of the ER and
  therefore is proportional to number of vibrations (varying as a Boltzmann factor).
  Turning on the radiative channel ($\beta^{rad}>0$), being it
  independent on temperature, we obtain a low temperature saturation of the UP PL and the $C$ ratio exhibits a plateau as experimentally
  found \cite{schouwink, ceccarelli}.
  The experimental data are well reproduced and stay between the simulations with
  $\beta^{rad}=0.05$ and $\beta^{rad}=0.1$.

  \begin{figure}
    \centering
    \includegraphics[width= 8.3cm]{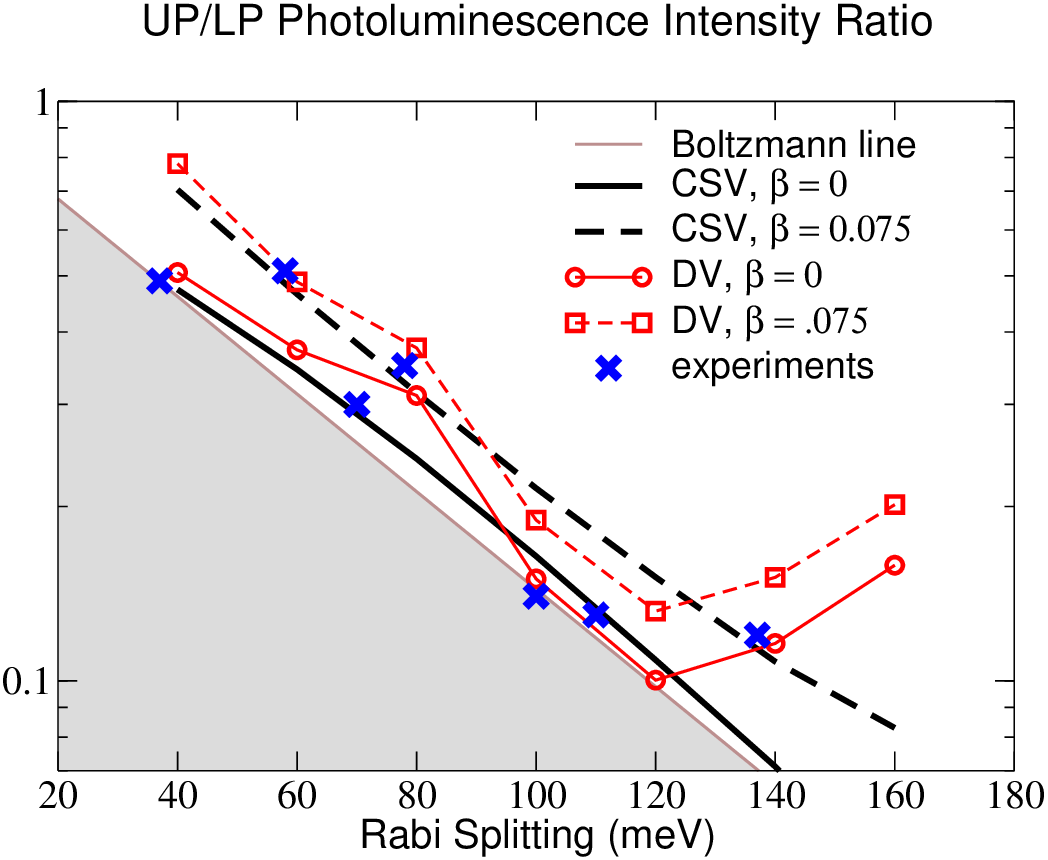}
    \caption{(Color online) The $C$ ratio as function of the Rabi splitting for microcavities with $\beta=0$ and $\beta=0.075$, for the models with
      continuous and discrete spectrum of vibrations.
      The calculation are compared with the experimental data in blue ``Xs''\cite{chovan}.}
    \label{fig:rabi}
  \end{figure}
  We also show in Fig.\ref{fig:rabi} the simulation of the $C$ ratio as a function of the Rabi
  splitting in comparison with experimental data from Ref.\cite{chovan}.
  Increasing the Rabi splitting value reduces the probability for a
  particle in the ER to be scattered on the UP branch, because such
  process would need the absorption of a vibration of greater energy
  from the thermal bath.
  The data therefore roughly follows a Boltzmann factor
  $e^{-\frac{\Delta}{2K_bT}}$, shown
  for comparison, where $\Delta/2$ corresponds, at resonance angle, to the energy
  separation from the ER bottom and the UP and LP states.
  With a continuous vibrational spectrum we observe the smooth
  Boltzmann behavior, while the inclusion of the radiative channel rigidly shifts the $C$ ratio toward higher values.
  The experimental data (blue X symbols) stay in between the
  simulation performed without the radiative channel $\beta=0$ and the simulation
  with the radiative process turned on with $\beta=0.075$.

  \subsection{Model with discrete vibrations}
  The temperature dependence of the $C$ ratio, shown in Fig.\ref{fig:temp} is largely unaffected by
  the introduction of the discrete spectrum of vibration and the same
  considerations as for the CSV case still apply.  
  If we instead consider the Rabi splitting dependence, the inclusion
  of discrete energy vibrations leads to the fluctuation of the $C$
  ratio and to a deviation from the smooth Boltzmann-like behavior.
  In particular the DV model reproduces the drop in the $C$ factor between $80$
  to $100$~meV, associated with $\Delta/2\approx40$~meV and the activation of the UP depletion towards
  the ER bottom with emission of a vibrational quantum and permits to
  obtain a better fit of experimental data.
  We believe
  that the agreement we obtain with the available data as show in Fig.\ref{fig:rabi} is of
  comparable quality to that obtained by Chovan et
  al.\cite{chovan}. 
  Of course, this is not to say that the phonon strong
  coupling approach they consider is immaterial.  
  It is plausible that also in J-aggregate microcavities exciton and molecular vibrations may interact strongly
  leading to non perturbative effects \cite{holmes,fontanesi,embriaco} such as, in particular,
  the formation of cavity polaritons with the concomitant participation of
  optical phonon, exciton and cavity photon modes, a phenomenon
  which has also been considered in bulk semiconductors (see the
  concept of "phonoriton" \cite{phonoriton}).
  It remains to be seen,
  however, whether there is as yet a compelling experimental evidence
  for the occurrence of such a phonon strong coupling in
  J-aggregate microcavities, the photoluminescence of which is here considered.

  Apart for the particular choice of the definite energy vibrations we made in the DV
  model ($E_{CSM}=25$~meV, $E_1=40$~meV, $E_2=75$~meV), we also tested different sets of optical vibrations in order to
  understand their effects on the $C$ ratio as a function of the Rabi
  splitting.
  In general it is not possible to completely distinguish the
  individual effects of each one of $E_{CSM}$, $E_1$, $E_2$, however
  we are able to identify some trends.

  \begin{figure}
    \centering
    \includegraphics[width= 8.3cm]{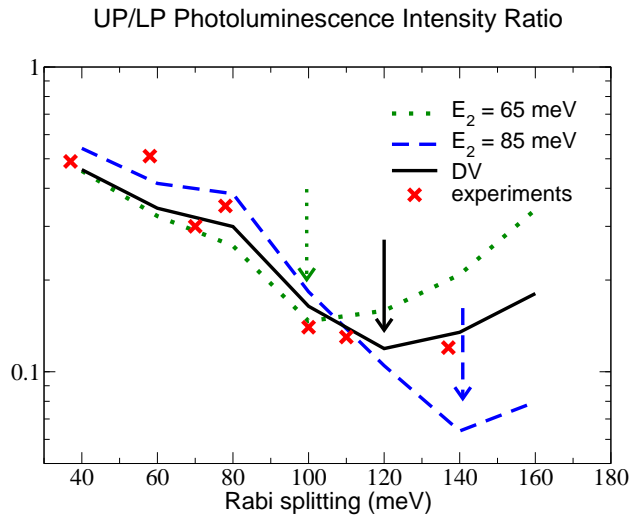}
    \caption{(Color online) $C$ ratio as function of the Rabi splitting for
      microcavities with $\beta=0$, for the DV model ($E_2=75$~meV) in
      continuous line, with $E_2$ changed
      in $65$~meV (dotted line), with $E_2=85$~meV (dashed line).
      The calculation are compared with the experimental data in blue ``Xs''\cite{chovan}.}
    \label{fig:rabic}
  \end{figure}
  Firstly we analyze the role of optical vibration $E_2$ by increasing it
  from $65$~meV to $85$~meV.
  As shown in Fig.\ref{fig:rabic} a minimum is observed at different
  $\Delta$ for different values of $E_2$.
  The motivation is that the ER to LP scattering rate is in
  resonance when $E_{ER}-E_{LP}\approx E_2$, with $E_{ER}$ averaged
  on the steady state population of the exciton reservoir.
  When the process is in full resonance we expect a minimum of the $C$
  factor, as it is evidenced in Fig.\ref{fig:rabic}.
  Following this rule we expect the minimum to be located in $\Delta_{min}=
  2(E_2-\varepsilon)$.
  Because of the different position of $\Delta_{min}$, when ER to LP
  scattering is maximum, for small Rabi splitting the $C$
  factor decrease with $E_2$, while we find the opposite behavior for $\Delta$
  greater than $100$~meV.
  As we will see, the occurrence of the $C$ minimum depends on the
  presence of higher energy vibrations active in relaxation processes,
  here not included.

  \begin{figure}
    \centering
    \includegraphics[width= 8.3cm]{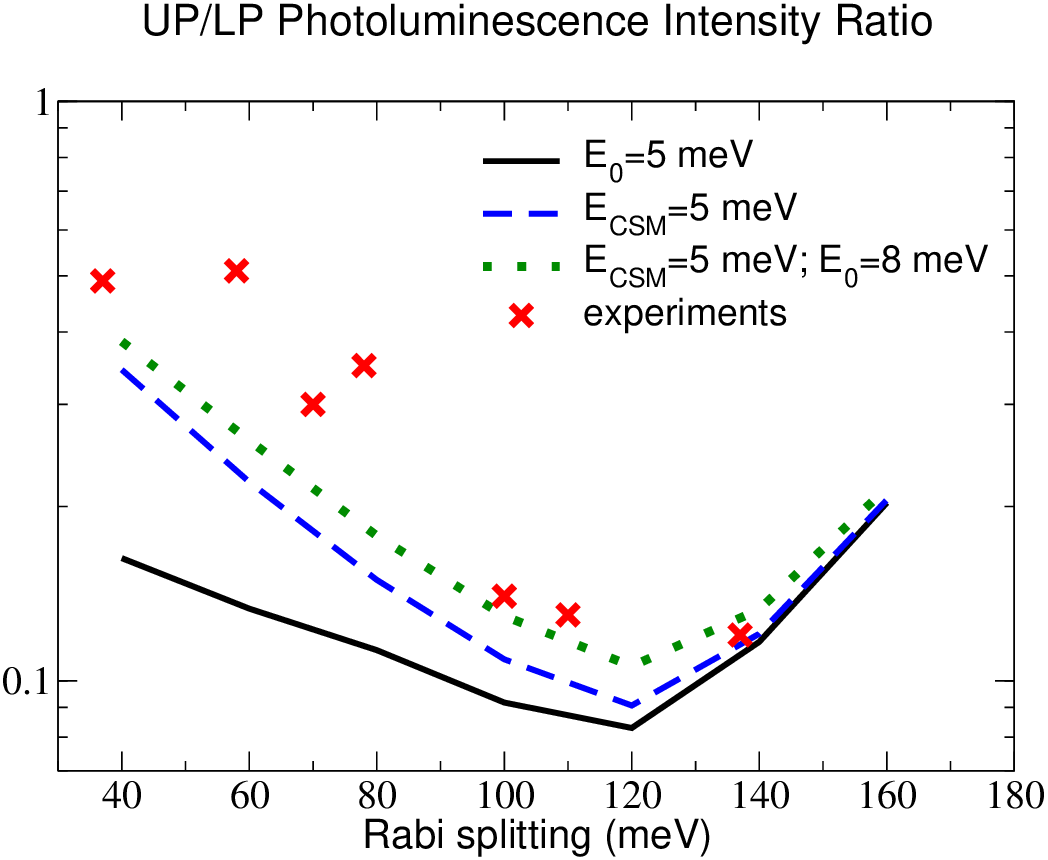}
    \caption{(Color online) $C$ ratio as function of the Rabi splitting for
      microcavities with $\beta=0$ and low energy vibrational spectrum
      modified with respect to the DV model. 
      Simulation with no CSV and a defined energy vibration $E_0=5$~meV
      (continuous line); with $E_{CSM}=5$~meV (dashed); with
      $E_{CSM}=5$~meV, plus a defined energy vibration $E_0=8$~meV (dotted).
      The calculation are compared with the experimental data in blue ``Xs''\cite{chovan}.}
    \label{fig:rabia}
  \end{figure} 
  We now focus on the role of the low energy vibrations.
  Initially we modify the DV model including the single vibration $E_0=5$~meV instead of
  the continuous spectrum of low energy vibrations.
  The result, in Fig.\ref{fig:rabia}, is a significant reduction of the $C$ factor for $\Delta$
  smaller than $120$~meV.
  We can understand this if we consider that UP population is mainly due to
  the ER to UP scattering, that takes place from steady state ER
  population with absorption of a molecular vibration, while the
  direct relaxation in UP from high energy excitons gives a minor contribution. 
  This process involving absorption of a vibrations is active at room
  temperature till around $E_0=K_bT$ (if the vibrational
  DOS is not zero), therefore it is significant for UP energies equal
  to the mean ER population energy ($\varepsilon$) plus $k_b T$, that
  corresponds to Rabi splitting values $\Delta<2(\varepsilon+K_bT)\approx 80$~meV. 
  Therefore a reduction in the active low energy DOS means a
  significant reduction of the ER to UP scattering and of the $C$ ratio for small $\Delta$.
  If we modify DV choosing $E_{CSM}=5$~meV we note that after $\Delta=
  2(\varepsilon+E_{CSM}) = 40$~meV we have a rapid decrease of
  the $C$ factor and a minimum, as predicted around $120$~meV.
  So it is clear that to recover the experimental data we need a series
  of active vibrations between $5$ and $25$ meV.

  Finally let's consider the effects of a third optical vibration
  $E_3>E_2$, in particular we set $E_3=120$~meV or $E_3=160$~meV with $g_3=0.3$.
  \begin{figure}
    \centering
    \includegraphics[width= 8.3cm]{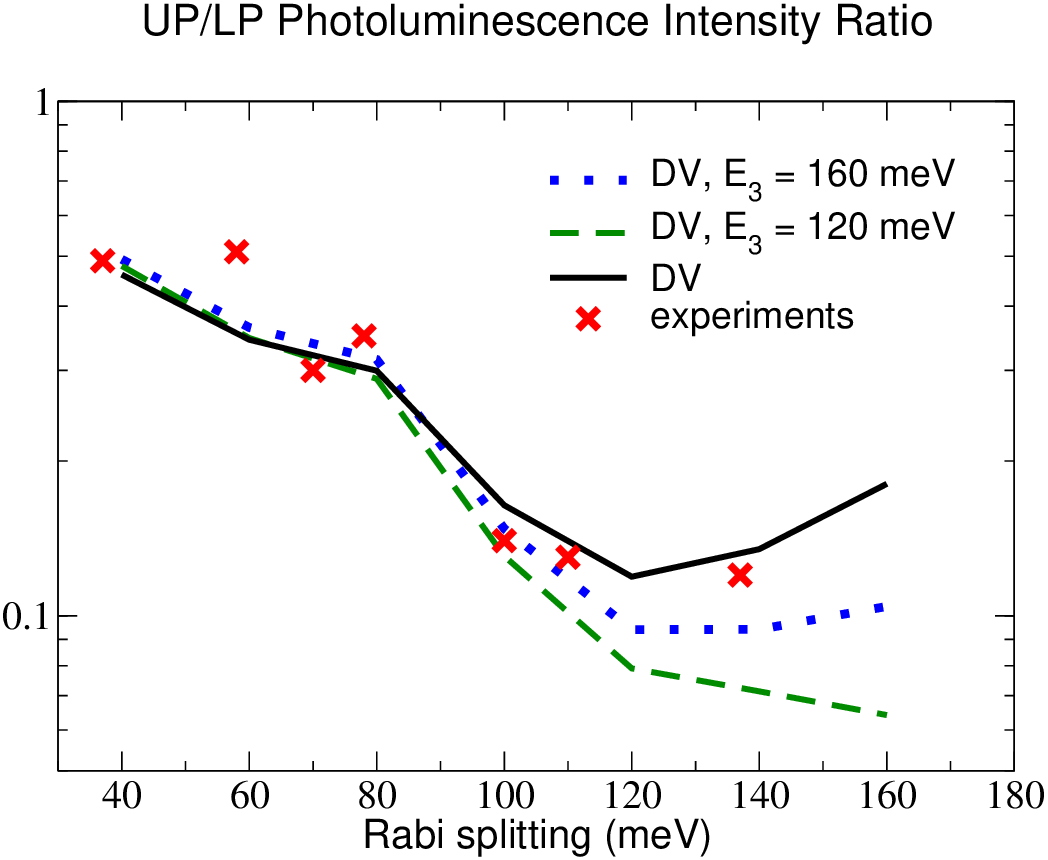}
    \caption{(Color online) $C$ ratio as function of the Rabi splitting for
      microcavities with $\beta=0$, for the DV model plus an high
      energy vibration $E_3=120$ (dotted), $160$~meV (dashed). 
      The calculation are compared with the experimental data in blue ``Xs''\cite{chovan}.}
    \label{fig:rabid}
  \end{figure}
  As can be seen in Fig.\ref{fig:rabid} the presence of $E_3$ has
  effects on $C$ only for large enough Rabi splitting, for which
  $\Delta/2 + \varepsilon$ approach $E_3$. 
  For $E_3=120$~meV, with $\Delta$ in the range considered, the
  non-monotonic behavior of $C$ vanishes correspondent to the fact
  that the LP is pumped from the ER almost thermalized population with
  an emission of a vibration $E_3$.
  If we increase $E_3$ to $160$~meV the ER to LP scattering become too
  much out of resonance to contribute and numerical data are nearer to
  the DV case.
  It must be noted that the possible presence of a $C$ minimum is conditioned to the
  absence of optical vibrations of higher enery $E_3$ active in
  relaxation processes.

  \section{Conclusions}
  In conclusion our model, based on semiclassical
  rate equations, explains sufficiently well the presently available
  experimental data on strongly coupled J-aggregate microcavities.
  In particular, the photoluminescence under
  non-resonant pumping follows the scattering of an electronic
  excitation from the ER bottom, where the population is
  accumulated, towards the polariton branches.
  Such scattering can either happen with the absorption/emission of
  vibrational quanta (with a continuous as well as discrete
  spectrum), process which is sensitive to temperature, or be due to
  pumping of polaritons by the spontaneous radiative decay of weakly coupled
  excitons.
  The interplay between the two processes determines both the
  specific temperature dependence of the polariton branches
  photoluminescence intensity ratio and its Rabi splitting
  dependence.

\end{document}